\documentclass[nofootinbib,showpacs,11pt]{revtex4}
\usepackage{latexsym}
\usepackage{amsmath}
\usepackage{amsthm}
\usepackage{fancyhdr}
\usepackage{graphicx}

\begin{document}

\title{Remarks on the complex branch points in $\pi N$ scattering
amplitude\\
and the multiple poles structure of resonances}

\author{Shin Nan Yang}
%\email{}
\affiliation{Department of Physics and Center for Theoretical Sciences, National Taiwan University, Taipei 10617, Taiwan}

\begin{abstract}

A simple heuristic argument to understand the existence of branch
points in the unphysical sheet for $\pi N$ scattering amplitude is
presented. It is based on a hypothesis that the singularity
structure of the $\pi N$ scattering amplitude is a smooth varying
function of the pion mass. We find that, in general, multiple poles
structure of a  resonance is a direct mathematical consequence when
additional Riemann surface is included in the study and the two-pole
structure found to correspond to the Roper resonance is a good
example.
\end{abstract}


\maketitle

%\section{Introduction}
Recently, there is a renewal of interest on the existence of complex
branch cut and their relevance in the partial wave analysis. It
arises, in large part, from the increasing focus on the properties
of the Roper and other resonances in $\pi N \,\,P_{11}$ channel
after the electro-excitation properties of the $\Delta(1232)$  has
been much studied \cite{Pascalutsa07}. Two   features in the
analysis of $P_{11}$ channel have received special attention.
Namely, the need to include a branch cut starting at the branch
point $m_\pi+M_\Delta$ in the complex plane because of the opening
of $\pi\Delta$ channel and consequently a two-pole structure was
found to correspond to the Roper resonance.

The inclusion of complex cuts in the pion-nucleon partial wave
analysis has a long history. This is because  pion is light and the
inelastic channel $\pi\pi N$ appears already at C.M. energy of
$W=1216$ MeV. The effect of the three-body channel of $\pi\pi N$ is
most often treated via coupled-channels approach where the
 quasi two-body channels like $\pi\Delta, \eta N$ and
$\rho N$ etc. are introduced  \cite{Cutkovsky79}. It generates
naturally branch cuts in the second sheet starting from the quasi
two-body inelastic threshold.  However, only the poles reached most
directly by analytic continuation from the real axis were looked for
in \cite{Cutkovsky79} and only one pole was found to correspond to
the Roper. The poles behind the complex branch cuts were first
studied in SAID's pion-nucleon partial-wave analysis \cite{Arndt85}.
It was then noticed that there are two poles associated with the
Roper resonance $P_{11}(1470)$ in the Riemann surface of the partial
wave amplitude. It led Cutkovsky and Wang \cite{Cutkovsky90} to
re-examine the previous analysis of \cite{Cutkovsky79} and confirmed
that indeed there are two and four nearly degenerate poles at 1470
and 1700 MeV, respectively, if poles in other Riemann sheets were
searched for.

In a recent study \cite{Suzuki10} by the EBAC (Excited Baryon
Analysis Center) group, it is demonstrated, via a dynamical
coupled-channels model (JLMS) \cite{Julia-Diaz07} that the two
almost degenerate poles near the $\pi\Delta$ threshold and the next
higher mass pole in the $\pi N \,\,\, P_{11}$ channel actually all
evolve from a single bare state through its coupling with $\pi N,
\,\,\eta N,$ and $\pi \pi N$ reaction channels. It confirms the
previous conjecture that the two-pole structure found in
\cite{Arndt85,Cutkovsky90} are indeed associated with the Roper
resonance. The $\pi\Delta$ complex branch cut is not considered in
the dynamical models of Dubna-Mainz-Taipei (DMT)
\cite{KY99,Hung01,Kamalov01,Kamalov01a} and Sato-Lee \cite{Sato96},
which were developed in the 1990's based on a dynamical approach
proposed in \cite{Yang85,Tanabe85}, even though $\Delta$ degree of
freedom is explicitly considered. This is why only one pole is found
to correspond to the Roper resonance in the DMT $\pi N$ model
\cite{Tiator10}.

\begin{figure}[h]
\begin{center}

\includegraphics[width=0.6\linewidth,angle=0]{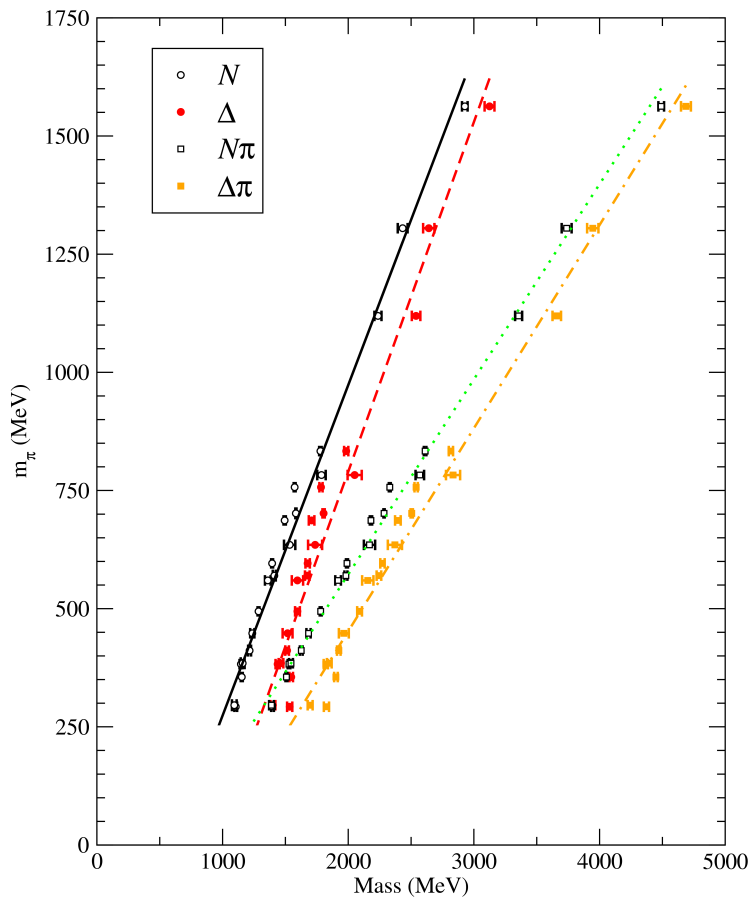}
\end{center}

\caption{Lattice QCD results for the evolution of the masses of the
nucleon (N), $\Delta(1232)$, $\pi N$, and $\pi\Delta$ with pion
mass. Data are from \cite{HSC11,LHP05,CSSM10,PACS09}.} \label{LQCD}
\end{figure}

The  branch point in the complex plane has been shown to exist using
only the general properties of the S matrix \cite{Ceci11} and
demonstrated to be important for  reliable extraction of resonance
parameters. In this short note, we present  a simple heuristic
argument to understand the existence of branch points in the
unphysical sheet for $\pi N$ scattering amplitude. It is based on a
hypothesis that the singularity structure of the $\pi N$ scattering
amplitude is a smooth varying function of the pion mass, or
equivalently the quark masses. We further point out that the
two-pole structure found to correspond to the Roper resonance is
actually a direct mathematical consequence when additional Riemann
surface is included in the study and will be a general feature for
all
resonances when multi-Riemann sheets are considered.\\

\noindent{\it Existence of complex branch cut
in $\pi N$ scattering}\\

\begin{figure}[h]
\begin{center}
\includegraphics[width=0.8\linewidth,angle=0]{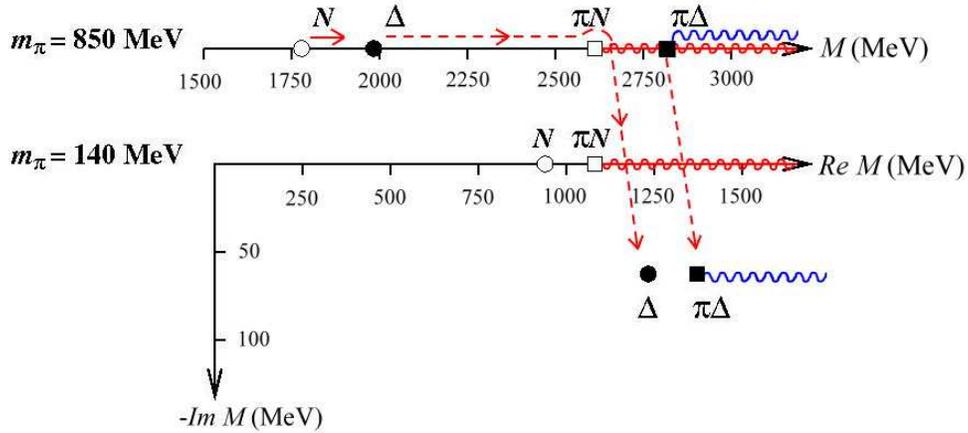}
\end{center}
\caption{Movement of the $\Delta$ pole and the $\pi\Delta$ branch
cut with change of pion mass in the LQCD results if variation w.r.t.
$m_\pi$ would be smooth.} \label{branch-cut}
\end{figure}

For simplicity, we illustrate our argument for the existence of
complex branch cut in $\pi N$ scattering by considering only $\pi N$
and $\pi\Delta$ channels. Fig. \ref{LQCD} shows the lattice QCD's
results for the masses of the nucleon (N), $\Delta(1232)$, $\pi N$,
and $\pi\Delta$, represented by open circles, solid circles, open
squares and filled squares, respectively, with pion mass varies from
around 1550 MeV down to about 300 MeV. The data are from
\cite{HSC11,LHP05,CSSM10,PACS09}. Details can be found in a recent
review \cite{Lin11}. The straight lines are shown to guide the eyes
only as it is well-known that linear extrapolation is not valid at
low pion mass region and chiral extrapolation is called for.  In the
large pion mass region, say, $m_\pi \ge 850$ MeV, one always has
$m_\pi + M_\Delta > m_\pi + M_N
> M_\Delta > M_N$ such that $\Delta$ is bound and stable.
Consequently, there are two branch cuts, denoted by the wiggly
lines, starting at $m_\pi + M_N$ (denoted by open square) and $m_\pi
+ M_\Delta$ (denoted by filled square) on the real axis,
respectively, as shown in the upper horizontal line, labeled with
$m_\pi=850$ MeV on the left, of Fig. \ref{branch-cut}.

It is further seen in Fig. \ref{LQCD} that as pion mass decreases,
both the nucleon and $\Delta$ masses decrease as well, but with
$M_N$ decreasing at a faster pace. In addition, the lattice data
show that both the open and solid circles representing nucleon and
$\Delta$ masses   move closer to $\pi N$ threshold as $m_\pi$ gets
smaller. This is indicated by the arrows on the solid and dashed
lines just above the horizontal line labeled with $m_\pi=850$ MeV in
Fig. \ref{branch-cut}, as pion mass decreases. Eventually lines
representing $M_\Delta$ and $m_\pi + M_N$ in Fig. \ref{LQCD} would
cross at around $m_\pi = 300$ MeV and the solid circle representing
$\Delta$ would move to the right of the open square which
corresponds to the branch point of $\pi N$ elastic cut. $\Delta$
would then

\noindent  become unstable and should begin moving into complex
plane, as shown by the dashed line connecting $\Delta$ on the upper
horizontal line and the $\Delta$ lying in the complex plane below
the lower horizontal line labeled with $m_\pi=140$ MeV on the left,
in Fig. \ref{branch-cut}. We have purposedly  aligned the two open
squares corresponding to the $\pi N$ elastic threshold obtained both
with $m_\pi=850$ and $m_\pi=140$ MeV to show more clearly how
$\Delta$ pole moves as pion mass is varied. The lattice calculation
is not sophisticated yet to calculate the width of an unstable
particle at present. In Fig. \ref{branch-cut}, an experimental value
of $\Gamma_\Delta=120$ MeV for the width of the $\Delta$ is assumed
for the  LQCD result if it will become  possible to calculate it on
the lattice.

The pole character of the $\Delta$ in the $\pi N$ scattering
amplitude should remain, as generally expected, unchanged after it
moves into complex plane, if the singularity structure of the S
matrix would vary smoothly with the pion mass. In the same token,
the branch point corresponding to the opening of $\pi\Delta$ would
also move into complex plane and its squared root character should
  be retained as well. Accordingly, there should exist a branch cut in
the complex plane starting from $m_\pi +M_\Delta$, which is complex
when the value of $m_\pi$ goes down to 140 MeV, as indicated in Fig.
\ref{branch-cut}.
\\

\noindent{\it Multiple poles structure of a resonance with two Riemann sheets}\\

As the mass of the Roper resonance lies close to the $\pi\Delta$
threshold, it is natural to expect that it is important to include
the $\pi\Delta$ branch cut in the extraction of the properties of
the Roper as demonstrated in \cite{Ceci11}. The inclusion of the
$\pi\Delta$ branch cut in the partial wave analysis in $P_{11}$ has
led to the conclusion \cite{Arndt85,Cutkovsky90,Suzuki10} that there
are two almost degenerate poles corresponding to the Roper, one on
the unphysical sheet directly reachable from the real axis and the
other located just behind the $\pi\Delta$ cut. We   demonstrate in
the followings that such a two-pole structure is actually a direct
mathematical consequence when there are two Riemann sheets to be
considered.

The structure of Riemann surface with two branch cuts of squared
root nature present depends on whether they appear in   product or
additive form, namely, like $\sqrt{(z-z_a)(z-z_b)}$ or
$\sqrt{z-z_a}+\sqrt{z-z_b}$. It is known \cite{churchill84} that in
the case of the product type like $\sqrt{(z-z_a)(z-z_b)}$,  only two
Riemann sheets are needed to make it a single-valued function. The
Riemann surface in this case  is often represented by drawing a cut
connecting the two branch points $z=z_a$ and $z_b$. However, if the
two branch points appear in additive form like
$\sqrt{z-z_a}+\sqrt{z-z_b}$, then four Riemann sheets are required
to make it   a single-valued function. Since the branch points in
the pion-nucleon scattering arise from the opening of inelastic
channels in the intermediate states, the branch points will appear
in additive manner. Accordingly, we will discuss only the case where
two branch cuts appear in additive form.

A simple example where we have a pole at $z=z_0$ in the presence of
two additive branch cuts, both of squared root nature, would be a
function of the following form,
\begin{equation}
(a).\hspace{0.5cm}
f_1(z)=\sqrt{z-z_a}+\sqrt{z-z_b}+\frac{g(z)}{z-z_0},
 \label{f1}
\end{equation}
where $g(z)$ is an arbitrary function. As mentioned in the above, we
need a Riemann surface consisting of four sheets to make $f_1(z)$
 a single-valued function, with two cuts starting from $z=z_a$ and
$z=z_b$. It is then a simple mathematical exercise to see that the
pole $z=z_0$ would appear in all four sheets. The residues at these
four poles would all be equal if $g(z)$ is an analytical function.

For a slightly more complicated case like,
\begin{equation}
(b).\hspace{0.5cm}
f_2(z)=\sqrt{z-z_a}+\sqrt{z-z_b}+\frac{h(z)}{\sqrt{z}-z_0},
 \label{f2}
\end{equation}
where $h(z)$ is an arbitrary function, the pole   now appears at
$z=z^2_0$. This could occur if $g(z)$ in Eq. (\ref{f1}) behaves like
$\sqrt{z}+z_0$ around $z_0$. At first sight, one would think that we
need eight Riemann  sheets to make $f_2(z)$ single-valued, since
there are now three branch points at $z=z_a, z_b$ and 0. However, a
closer analysis \cite{Cheng11} would lead to the conclusion
  that only four Riemann sheets are needed and the pole
$z=z^2_0$ would appear only in two sheets. The residues at these two
poles would be, in general, different, even if $h(z)$ is an
analytical function.

In the DMT $\pi N$ model, the   contribution of a resonance $R$ to
the t-matrix takes the form \cite{Chen07},
\begin{equation}
t_{\pi N}^{R}(E)=\frac{\bar h_{\pi R}(E) h_{\pi
R}^{(0)}}{E-M_{R}^{(0)}(E) -\Sigma_{R}(E)},\label{tmatrix}
\end{equation}
where $h_{\pi R}^{(0)}$ and $\bar h_{\pi R}(E)$ describe the bare
and dressed vertices of $\pi N\rightarrow R$, and $M_{R}^{(0)}$ and
$\Sigma(E)$ denote the bare mass and the complex self-energy of
resonance $R$, respectively. Both $\bar h_{\pi R}(E)$ and
$\Sigma(E)$ receive dressing from the background potential which
would contain  information related to the branch cuts associated
with the background mechanisms. Consequently, the function $g(z)$ or
$h(z)$ could contain factors pertaining to the squared root branch
cuts like $\sqrt{z-z_a}$ and $\sqrt{z-z_b}$. Nevertheless, the
conclusions we have obtained for cases (a) and (b) concerning the
multiplicity of the resonance pole remain unchanged except that the
residues will now all be different, even in the case of (a).

The conclusions we obtain in the above concerning the multiplicity
of the resonance pole are consistent with the results obtained in
\cite{Cutkovsky90,Arndt85,Suzuki10} from $\pi N \,\,\, P_{11}$
partial wave analysis and the subsequent analytical continuation. In
particular, the findings in \cite{Cutkovsky90} that there were two
sets of two almost degenerate poles and one set of four almost
degenerate poles,   all with different residues can be readily
understood by our simple mathematical analysis.

In summary, we have presented a simple heuristic argument to
understand the existence of branch points in the unphysical sheet
for $\pi N$ scattering amplitude. The reasoning is based on a
hypothesis that the singularity structure of the $\pi N$ scattering
amplitude is a smooth varying function of the pion mass. We find
that the two-pole structure found to correspond to the Roper
resonance is actually a direct mathematical consequence when
additional Riemann surface is included in the study. In fact, our
analysis indicates that there are always multiple poles, either two
or four, to be associated with a resonance. The physical
significance of the poles lying in sheets not directly reachable
from the physical sheet remains to be further investigated and
established.

\section*{Acknowledgment}
I acknowledge gratefully the beneficial discussions  with Prof.
Kuo-Shung Cheng. I am also indebted to Drs. Huey-Wen Lin and Alvin
Kiswandhi for help in collecting  LQCD data and the plotting of the
figures, respectively. This work is supported in part by the
National Science Council of ROC (Taiwan) under grant No.
NSC100-2112-M002-012. \\

\end{document}